\documentclass[%
 aip,
 amsmath,amssymb,
 reprint,%
]{revtex4-1}

\usepackage{xcolor} % NOTE use command \revision_yourinitials to make revision in a colour you define
\definecolor{Red}{rgb}{1,0,0}
\definecolor{Orange}{rgb}{0.84,0.36,0.03}
\definecolor{Pink}{rgb}{1,0,1}
\definecolor{Blue}{HTML}{0069b4}
\definecolor{Black}{rgb}{0,0,0}
\newcommand{\revisionSam}[1]{\textcolor{Black}{{#1}}}

\newcommand{\revisionMatthias}[1]{\textcolor{Black}{{#1}}}
\usepackage{graphicx}% Include figure files
\usepackage{dcolumn}% Align table columns on decimal point
\usepackage{bm}% bold math
\usepackage{placeins}%for float barrier
\usepackage[utf8]{inputenc}
\usepackage[T1]{fontenc}
\usepackage{mathptmx}
\usepackage{etoolbox}

\definecolor{PPLN_Yellow}{RGB}{230, 230, 0}

\makeatletter
\def\@email#1#2{%
 \endgroup
 \patchcmd{\titleblock@produce}
  {\frontmatter@RRAPformat}
  {\frontmatter@RRAPformat{\produce@RRAP{*#1\href{mailto:#2}{#2}}}\frontmatter@RRAPformat}
  {}{}
}%
\makeatother

\begin{document}

\preprint{AIP/123-QED}

\title[]{Depth resolution in piezoresponse force microscopy}

%\title[PFM on buried domain walls]{Seeing what's below: Detection sensitivity of piezoelectric response force microscopy for buried ferroelectric domain walls}
% Force line breaks with \\
\author{Matthias Roeper}
\affiliation{TU Dresden, Institute of Applied Physics, Nöthnitzer Strasse 61, 01187 Dresden, Germany}

\author{Samuel D. Seddon}
\affiliation{TU Dresden, Institute of Applied Physics, Nöthnitzer Strasse 61, 01187 Dresden, Germany}

\author{Zeeshan H. Amber}
\affiliation{TU Dresden, Institute of Applied Physics, Nöthnitzer Strasse 61, 01187 Dresden, Germany}

\author{Michael Rüsing}
\affiliation{TU Dresden, Institute of Applied Physics, Nöthnitzer Strasse 61, 01187 Dresden, Germany}
\affiliation{Paderborn University, Integrated Quantum Optics, Institute for Photonic Quantum Systems (PhoQS), Warburger Str. 100, 33098 Paderborn, Germany}

\author{Lukas M. Eng}
\affiliation{TU Dresden, Institute of Applied Physics, Nöthnitzer Strasse 61, 01187 Dresden, Germany}
\affiliation{$\text{ct.qmat: Dresden-Würzburg Cluster of Excellence—EXC 2147, TU Dresden, 01062 Dresden,~Germany}$}

\email{Correspondence: lukas.eng@tu-dresden.de}

\date{\today}% It is always \today, today,

\begin{abstract}
Piezoresponse Force Microscopy (PFM) is one of the most widespread methods for investigating and visualizing ferroelectric domain structures down to the nanometer length scale. PFM makes use of the direct coupling of the piezoelectric response to the crystal lattice, and hence is most often applied to spatially map the 3-dimensional (3D) near-surface domain distribution of any polar or ferroic sample. Nonetheless, since most samples investigated by PFM are at least semiconducting or fully insulating, the electric ac field emerging from the conductive scanning force microscopy (SFM) tip, penetrates the sample, and hence may also couple to polar features that are deeply buried into the bulk of the sample under investigation. Thus, in the work presented here, we experimentally and theoretically explore the contrast and depth resolution capabilities of PFM, by analyzing the dependence of several key parameters. These key parameters include the depth of the buried feature, i.e. here a domain wall (DW), as well as PFM-relevant technical parameters such as the tip radius, the PFM drive voltage and frequency, and the signal-to-noise ratio. The theoretical predictions are experimentally verified using x-cut periodically-poled lithium niobate single crystals that are specially prepared into wedge-shaped samples, in order to allow the buried feature, here the DW, to be `positioned' at any depth into the bulk. This inspection essentially contributes to the fundamental understanding in PFM contrast analysis, and to the reconstruction of 3D domain structures down to a 1-$\mu$m-penetration depth into the sample.
\end{abstract}

\begin{keywords} {Piezoresponse Force Microscopy, PFM, depth resolution, periodically-poled lithium niobate, single crystal, PPLN, x-cut, nondestructive PFM.}
\end{keywords}

\maketitle

\section{\label{sec:Introduction}Introduction}

Piezoresponse Force Microscopy (PFM) is a contact-mode scanning force microscopy (SFM) technique that allows analyzing and visualizing ferroelectric domain structures \cite{GoldStandard2019,diss_haussmann,pfm_soergel,https://doi.org/10.1002/adma.202203449, https://doi.org/10.1038/s41467-019-10664-5,Seddon_2024} by making use of the inverse piezoelectric effect \cite{Gruverman2019}. In brief, an ac voltage is applied to the conductive SFM tip that acts as the top electrode, hence provoking the ferroelectric (FE) or polar sample under inspection to locally contract, expand, or shear  \cite{Gruverman2019,pfm_soergel,Eng1999}. For PFM, the tip stays in firm contact to the sample surface, thus directly translating any sample motion into a net 3D cantilever deflection that then is easily evaluated and monitored \cite{Gruverman2019,pfm_soergel}. Notably, the mechanical response of the cantilever depends on the orientation and magnitude of the piezoelectric sample tensor with respect to both the applied electric field $\vec{E}$ and the cantilever orientation \cite{diss_haussmann,Gruverman2019,pfm_soergel}. In a first order approximation \cite{Jungk2006}, the local piezoelectric response is directly proportional to the product of the local electric field $\vec{E}$ and the piezoelectric tensor element ${d_{ij}}$. Hence, when mapping the piezoelectric response, i.e. its magnitude and phase is measured via Lock-In-Amplifier (LIA) demodulation, which allows to evaluate and reconstruct the effective 3D orientation of ferroelectric domains as well as their 3D polarization \cite{Eng1999,Rodriguez2004,Kalinin2006}.

There is a strong interest in non-destructive imaging of domains, such as recently demonstrated by tomographic scanning electron microscopy\cite{He2024}. PFM is typically used to analyse the ferroelectric domain structure closest to the sample surface, achieving a superb lateral resolution down to only a few nanometer. Nonetheless, PFM may also be applied to probe into the depth of a sample, hence potentially revealing a clear contrast of buried nanostructures that are not `visible' at the sample surface. One such approach is tomographic PFM \cite{TAFM2019,Roede2019} following alternating PFM imaging with incremental sample removal; although resulting in a highly resolved 3D reconstruction of the FE polarization, that method is heavily destructive. 

In contrast, any feature that lies below the surface and is susceptible to PFM imaging, can also be measured non-destructively, since the electric field emanating from the spherical, conductive SFM tip converges to zero at an infinite depth, only. PFM thus, in principle, can provide depth information as well, provided some critical parameters in the PFM performance are known. \revisionSam{\citet{Lu2002}} \revisionMatthias{and others \cite{Eng2004} have reported a PFM depth dependence of the out-of-plane piezoresponse signal of up to 240~nm in thin-film ferroelectrics. \citet{depth_pfm} considered this by analyzing the PFM signal from buried spike domains created via UV laser-induced poling into single crystalline lithium niobate (LNO)}, demonstrating that domains buried even below 1 $\mu$m into the LNO bulk are still impacting the PFM amplitude. 

\begin{figure}[t]
	\centering
	\includegraphics[width=1\linewidth]{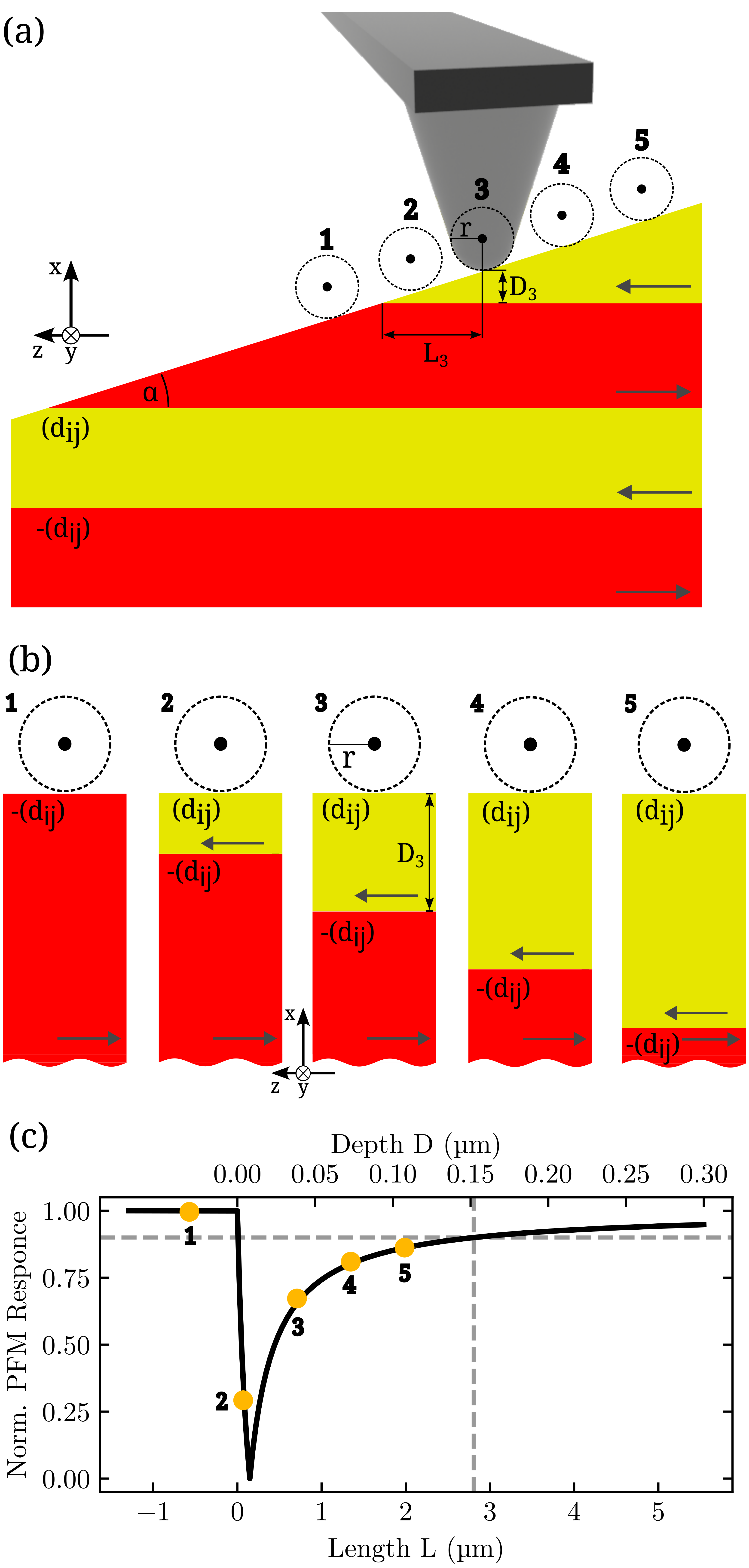}
	\caption{\revisionMatthias{(a) Sketch of the periodically-poled Lithium Niobate (PPLN) sample, polished at the x-face at an angle of $\alpha$ = 3.199 $\pm$ 0.012$^{\circ}$, creating shallow domains at the x-surface. (b) Sample volume beneath the tip for each position 1 to 5 as labelled in (a). This geometry was also used for calculating the thickness depend PFM signal in our theoretical model, with +z and -z domains shown in red and yellow, respectively. (c) Simulated relative PFM signal amplitude as a function of scan length $L$. The horizontal (x) axis is plotted in two versions, with the top axes revealing the corresponding depth D into the sample (D = 0 corresponds to the start of a new domain), and the bottom x-axis denoting the respective scan position L along the wedge surface.}}
	\label{fig:fig1}
\end{figure}

In the work presented here, this PFM probing depth is explicitly examined, both theoretically and experimentally, on a periodically-poled lithium niobate (PPLN) sample polished into a shallow-angled ($\alpha \sim 3$°) x-cut wedge to create well-defined domain transitions at a defined depth into the crystal [see Fig. \ref{fig:fig1}(a)]. The experimental setup parameters are systematically analyzed, such as the PFM drive voltage and frequency, the SFM tip force set point, the tip-radius, as well as the impact of the signal-to-noise ratio (SNR) on the depth resolution. Furthermore, it is shown how to accurately adjust both amplitude and phase whenever larger dc offsets impact the PFM signal.

\section{Methodology}
PFM measurements are performed on a Park NX10 SFM using the two internal LIAs. \revisionSam{Typical scan parameters used in this study are an applied ac voltage amplitude of +10~$V$ at an oscillation frequency of $\sim$ 80 $kHz$, and a cantilever set-point force of $\sim 950\  nN$. In the instance of varied scan parameters, these values are used whilst a singular parameter is varied.} Solid platinum tips manufactured by Rocky Mountain Nanotechnology, LLC of the type 25Pt300B, with a spring constant of 18 $Nm^{-1}$ are used for the majority of the measurements presented here. For tip-radius-dependent measurements, spherical shaped, high-density diamond-like carbon tips from nanotools GmbH of the types
biosphere B20-FM (spherical tip with radius $r_{tip} =$ (20~$\pm$~5)~$nm$),
biosphere B30-FM with $r_{tip} =$ (30~$\pm$~5)~$nm$,
biosphere B50-FM with $r_{tip} =$ (50~$\pm$~5)~$nm$,
biosphere B100-FM with $r_{tip} =$ (100~$\pm$~10)~$nm$, and
biosphere B150-FM with $r_{tip} =$ (150~$\pm$~10)~$nm$ are used. 
While these tips are not as conductive as the full metal tips, their conductivity is \revisionSam{sufficient to allow} for PFM measurements. If not otherwise noted, for the experiments reported here, it is the lateral PFM signal (in-plane signal) that was recorded (since the FE polarization points along the ±z-direction, i.e. in-plane for the x-cut LiNbO$_3$ sample). To do so, the cantilever is mounted with its long axis aligned in parallel to the y-direction of the crystal (see Fig. 1(a) for clarity). In this configuration a strong torsional motion of the cantilever is achieved through coupling to the $d_{15}$ tensor element of lithium niobate, which translates the vertical x-oriented electric field $E_x$ into an in-plane shearing motion in the yz-plane [see Fig. 1(a)].

Measurements are performed on a commercially available periodically-poled 5\%-MgO-doped lithium niobate (PPLN) single crystal having a periodic spacing $\Lambda =$ 16~$\mu$m and a duty cycle of nearly 50\%. Lithium niobate (LNO) is a uniaxial FE that, hence, allows the FE polarization to align along the +z and -z axes, only. For PPLN, the polarization hence alternately switches between these two orientations, resulting in a simple arrangement of $180^{\circ}$ domain walls (DWs) \cite{LN_Weis_1985,LN_Gopalan_2007} as depicted in Fig. 1(a). In fact, the DWs may be viewed as parallel sheets stacked along the x-axes. Now, to create buried domains and DWs, the PPLN sample is flipped onto its x-face, and then firmly embedded in epoxy resin, as shown in the supplement. The sample is then polished under a shallow angle \cite{SHG2021,SHG2022} [see Fig. 1(a)] of $\alpha$ = 3.199 $\pm$ 0.012$^{\circ}$ as determined with a confocal laser scanning microscope (Olympus LEXT, 405 nm excitation). As a result, the effective area of individual domains is enlarged by roughly a factor of 18 = $ \frac{1}{tan(\alpha)}$, hence allowing to allocate surface spots within the yz-plane with a defined x-depth to the next buried DW, with ultra-high precision. 

%As discussed in the supplement, due to cutting imperfections, this wedge angle does not necessarily represent the angle between the surface of the crystal and the planes of the domain walls. However, as shown in the supplement based, with an estimate of this error to be 10\%, which alike translates into an uncertainty of 10\% in calculated scaling of the depth axis. 

\FloatBarrier
\section{Results and discussion}
A schematic of the experiment as performed here, is depicted in Fig. \ref{fig:fig1}(a), showing the PFM tip (modelled as a sphere) being raster scanned along a gently inclined crystal surface, with the five illustrative points labelled in the schematic. As the ferroelectric crystal is periodically poled, the inclined polishing angle creates a wedged domain pattern, with the thickness of one such selected in-plane domain gradually increasing when scanning from left to right (i.e. from labels 1 to 5). Fig \ref{fig:fig1}(b) displays the local scenario for all these five selected sample areas separately. With the tip sitting at position 1, the "red" domain is the only contributor to the PFM signal, with its polarization pointing along the -z direction; only a negligible contribution from the next "yellow" domain is expected. As the tip gradually scans to positions 2 up to position 5, the tip passes a domain wall and the contribution of the "yellow" domain (polarized along the +z direction) gradually increases; notably, the poled red region becomes more and more buried and contributes less to the overall PFM signal. The key assumption made here is that all interfaces, i.e. the sample surface and any buried domain wall, are aligned quasi parallel, which significantly facilitates the modeling approach later. This assumption is well justified for the shallow polishing angle of $\alpha$ = 3.199 $\pm$ 0.012$^{\circ}$. 

Finally, Fig. \ref{fig:fig1}(c) plots the modelled and normalized PFM response at each tip stage 1-5, with the tip positions 1 to 5 being clearly indicated along the response curve. Qualitatively, the response is maximized whenever one type of domain is present only, as is for position `1'. %Notably, for the tip sitting close to or at the domain wall, the overall PFM response signal almost cancels (position `2'). 
Notably, for the tip sitting close to or at the position where the thickness of the "yellow" domain is equal to the tip radius, the overall PFM response signal almost cancels (position `2'). For tip positions `3' to `5', the new domain (here yellow) starts to dominate, with the buried (red) domain contribution becoming increasingly less significant.

\subsection{Modeling}

%Theoretical modeling of the experiments described above.
%equations, data, theoretical analysis.

The PFM response and hence the resulting cantilever deflection heavily depends on the local deformation of the crystal as a function of applied electric field $\vec{E}$. The crystal deformation is described by the strain tensor $\hat{\epsilon}$ that has 6 independent components $d_{ij}$ associated to uniaxial and shear strain. For piezoelectric materials, $\hat{\epsilon}$ is proportional to $\vec{E}$ as:
\begin{equation}\label{eq_epsilon}
	\mathbf{\hat{\epsilon}} = d_{ij}^\intercal \cdot \vec{E},
\end{equation} 
where $d_{ij}$ is the materials inverse piezoelectric tensor \cite{diss_haussmann}. The resulting motion of a cantilever in contact to a given domain structure can be calculated by starting from Eq.~(1) above \cite{Kalinin2002,Kalinin2004,Scrymgeour2005}. Here, the ferroelectric domain structure determines the spatial distribution and (local) numerical values of the piezoelectric tensor, whilst the local electric field is determined by the tip's location and shape with respect to the placement on the sample surface. The impact on the tip motion along different axes can be calculated by determining the total displacement at the tip's point of contact via integrating the local strain tensor component over the complete structure, weighed by the local electric field. For more accurate results further aspects like the elastic properties of the sample or the tip-sample-interaction need to be taken into account \cite{Kalinin2004}. 

Generally, such calculations are performed numerically using Finite-Element-Modeling (FEM) approaches \cite{Kalinin2002,Kalinin2004,Scrymgeour2005}. Here, however, we decided to perform an analytical analysis of a simplified structure, which - as will be shown below - provides already an accurate description. The following assumptions and simplifications are made to describe the problem analytically: \\ 
The sample is assumed to extend infinitely in the yz-plane, and is treated as a plan parallel layer of thickness $D$ for the $+z$ domain that sits on an infinitely thick domain with $-z$ orientation, as shown in Fig. 1(b). This manifests a well justified approximation due to the small wedge angle $\alpha$, and has already successfully been applied to comparable optical simulations \cite{SHG2021,SHG2022}. Since LNO is an uniaxial ferroelectric, the tensor elements in the $+z$ and $-z$ simply invert from $d_{ij}$ to $-d_{ij}$. Furthermore, the tip is assumed to be of spherical shape (tip radius $r_{tip}$, effectively representing a point charge at the sphere's center) to account for the tip's emanating electric field $\vec{E}$. Then, the electric field inside the sample can be easily described through image charges \cite{Bartelmann2018}. Hence, $\vec{E}$ follows as \cite{depth_pfm, Bartelmann2018}:  

\begin{equation}\label{eq_vecE}
	\vec{E} \propto \frac{2\cdot Q}{\epsilon_r + 1}\cdot\frac{\vec{R}}{\mid\vec{R}\mid^3}.
\end{equation} \\

In Eq.~(\ref{eq_vecE}), $Q$ is the magnitude of the point charge located in the tip center, $\epsilon_r$ is the relative permittivity of the material, while $\vec{R}$ is a radial vector originating at the point charge\cite{Bartelmann2018}. %, while $\vec{e_x}$ is a unit vector along the $x$ direction\cite{Bartelmann2018}.
The electric field distribution behaves rotationally symmetric (for small $\alpha$), meaning, that to a first approximation, any strain contribution outside this axis pairwise cancels. Hence, only the $x$-component of the electrical field $E_x$ \revisionMatthias{along $\vec{R}=\left(R_x,0,0\right)^\intercal$} needs to be calculated, simplifying Eq.~(\ref{eq_vecE}) to:
\begin{equation}\label{eq_1/z}
	E_x \propto \frac{1}{R_x^2}, \ \ \ \text{for} \ R_x > r_{tip}.
\end{equation}

Thus, the magnitude of the electric field $E_x$ within the sample is inversely proportional to $R_x^2$.
%the depth $D$ within the sample where the actual PFM signal stems from. %Note that $D = x$ for $r_{tip} \rightarrow 0$ as frequently used. 
With these assumptions, the PFM amplitude $A$ can be readily calculated by integrating the piezoelectric response over the relevant volume, i.e. the electric-field-induced strain according to Eq.~(\ref{eq_epsilon}). For the case here, the only non-zero field-component is along the crystal's x-axis, and primarily owed to the $d_{15}$ component. Then, for a given $+z$ top domain of thickness $D$, the PFM amplitude $A(D)$ results as: 

\begin{multline}\label{eqn:division2}
     A(D) \propto \left[ \int_{r_{tip}}^{D+r_{tip}} d_{15} E_x(R_x) dR_x  \right. \\ + \left.\int_{D+r_{tip}}^{\infty} (-d_{15}) E_x(R_x) dR_x\right], 
\end{multline}\\
where $(d_{15})$ and $(-d_{15})$ describe the addressed tensor elements in the $+z$ and $-z$ domain, respectively. Even when other tensor elements $d_{ij}$ are addressed, e.g. leading for example to an out-of-plane PFM signal, the same proportionality for the PFM amplitude signal holds, because, in any case, the piezoelectric tensor elements only change sign between domains of opposing polarization direction, at least for LNO. Therefore, the same calculus can be applied accordingly. The above calculation may also easily be expanded to describe hetero-structures of layered FEs or other DW types common to the ferroic crystal family, by using the appropriate magnitude and relation of the correct tensor elements.

Scanning along the z-direction on the wedged PPLN crystal hence produces plots of the integrated PFM amplitude $A(D)$ as displayed in Fig.~1(c). Here, $A(D)$ sharply decreases at $D = 0$, where the $-z$-oriented layer is added. A clear minimum is observed in $A(D)$ for $D = r_{tip}$ where the two weighted PFM responses from the
$+z$ and $-z$ layer cancel. Scanning beyond that point reveals the PFM signal $A(D)$ to recover hyperbolically due to the $D^{-1}$ dependence of an infinitely-thick domain. \\

To extract an easy-to-compare parameter between different experiments and simulations, it is convenient to define the $90\%$ depth $d_{90\%}$, which is the depth at which the PFM amplitude
\begin{equation}
	A(D_{90\%}) = 90\% \cdot A_{max}
\end{equation} \\ 
has recovered to 90\% of the maximum amplitude at infinite thickness $D$, similar to \citet{depth_pfm}. Solving this equation for the 90\%-depth yields:
\begin{equation}
	D_{90\%} = 19 \cdot r_{tip},
\end{equation} 
which is dependent on the tip-radius and a constant value, only. This means that with larger tip radii $r_{tip}$, domain transitions at proportionally larger depths can be detected. Conversely, the electric field emerging from a increasingly larger tip decreases at a slower rate. Nonetheless, a larger tip radius $r_{tip}$ also results in loosing the lateral resolution, hence causing a trade-off between depth and lateral resolution. 
\begin{figure}
	\centering
	\includegraphics[width=0.9\linewidth]{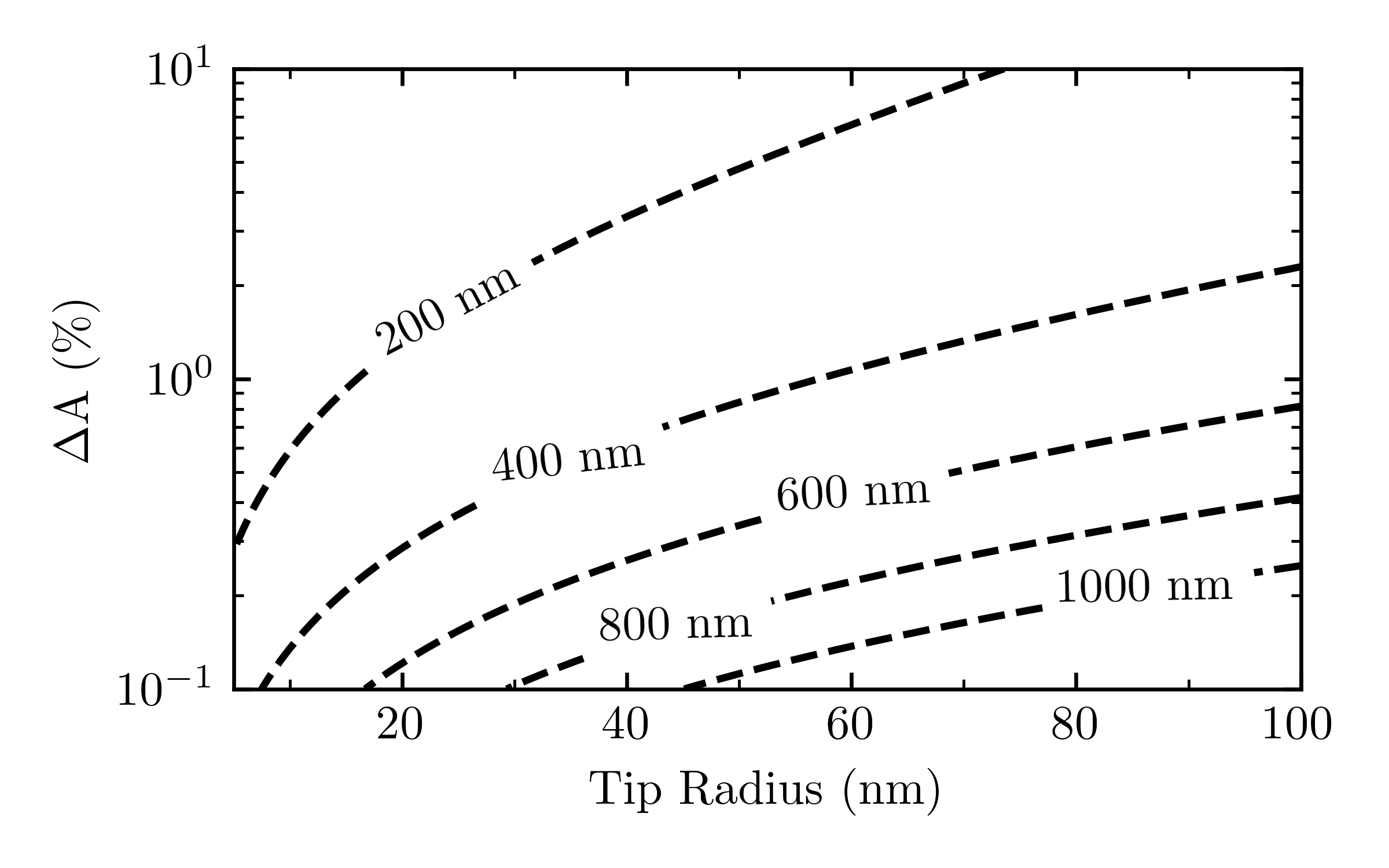}
	\caption{\revisionMatthias{Maximum possible depth (dashed lines) to achieve a depth resolution of 10~nm as a function of relative signal resolution ("noise level") $\Delta A$ and tip radius $r_{tip}$}.}
	\label{fig:fig2}
\end{figure}

The depth resolution, i.e. the error $\Delta D$ at a given depth $D$ that can be achieved, depends on the amplitude noise level $\Delta A$. When normalizing the recorded PFM amplitude $A$ to the maximum possible PFM amplitude $A_{max}$, the relative amplitude $A_{rel}(D)$ is obtained via the previous integrals as: 
\begin{equation}\label{eqn:arel}
A_{rel}(D) = 
	\left|1-\frac{2\cdot r_{tip}}{D+r_{tip}} \right|.
\end{equation}

In order to examine how strongly $\Delta D$ varies for an extracted depth $D$, the difference $\Delta A$ between two recorded amplitudes as resulting from the difference $\Delta D$ in the extracted corresponding depths is considered. This results in: 
\begin{equation}
    \Delta A_{rel} = A_{rel}(D \pm \Delta D) - A_{rel}(D).
\end{equation}

When combining with the definition of $A_{rel}(D)$ in Eq.~(\ref{eqn:arel}), the difference in recorded depths $\Delta D$ can be written as: 
\begin{equation}
\Delta D = \frac{2 \cdot r_{tip} }{\frac{2 \cdot r_{tip} }{(D + r_{tip} )}- \Delta A_{rel.}}   - r_{tip}  - D.
\end{equation}

The uncertainty in depth $\Delta D$ is the key value in order to determine, up to which depth $D$ PFM measurements are still feasible. A graph exploring the possible probing depths for a given resolution of $\Delta D = 10$~nm as a function of relative amplitude noise level $\Delta A$ and tip radius $r_{tip}$, is depicted in Fig. \ref{fig:fig2}. For example, for a given relative amplitude noise level of $\Delta A = 1~\%$, PFM measurements as deep as 400~nm are possible with a PFM tip of radius $r_{tip}=$ 60~nm, while this depth resolution shrinks to 200~nm for a tip radius $r_{tip}$ of 20~nm.

\subsection{Measurement procedure}

\begin{figure}
	\centering
	\includegraphics[width=0.9\linewidth]{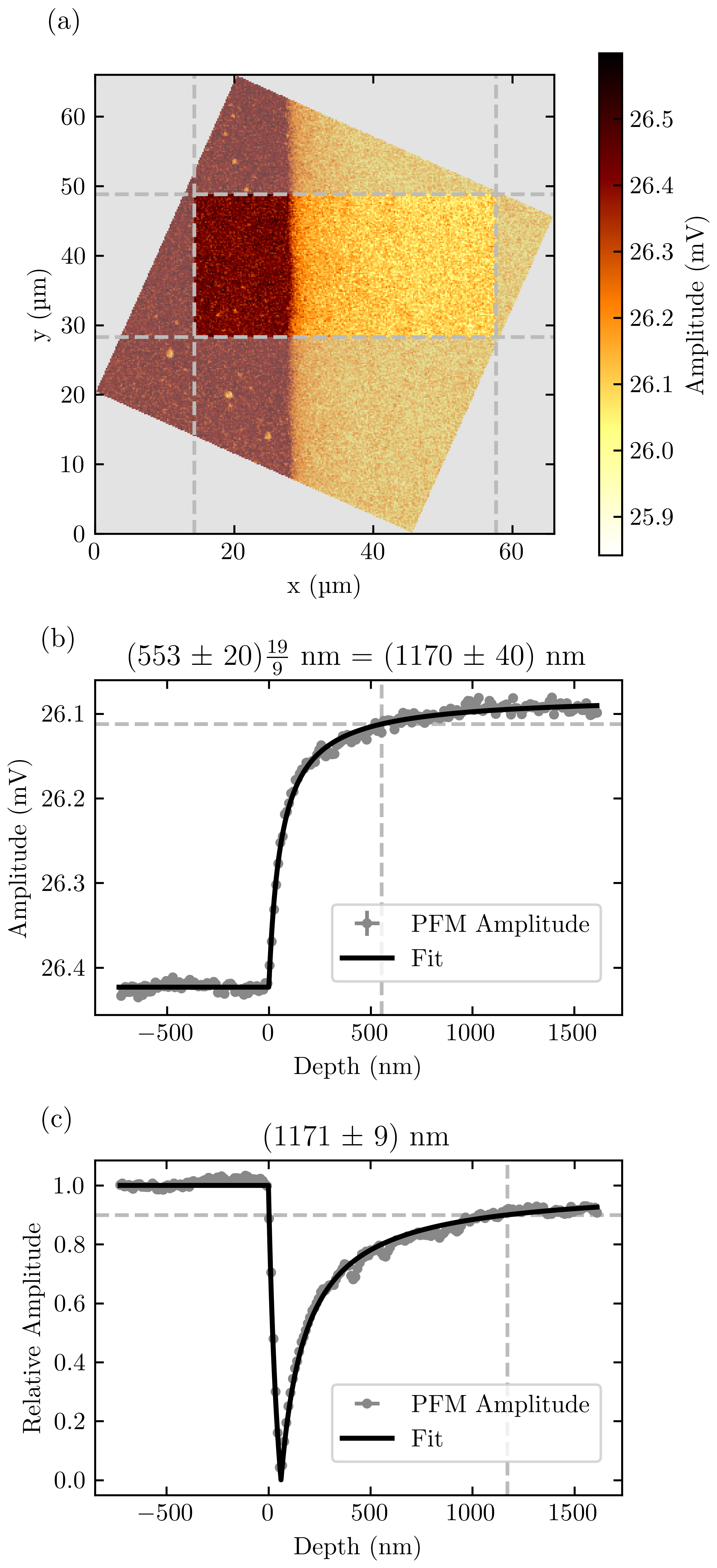}
	\caption{(a) A typical PFM amplitude image between two ferroelectric domains. The line scan extraction was averaged over the unshaded area. (b) Averaged line scan showing the raw data line scan, fitted to extract a 90\% depth of 1170 nm, after adjustment using the $\frac{19}{9}$ scaling factor. (c) Transformed data also indicating a 90\% depth of $\sim$ 1170 nm as fitted with the two-domain model.}
	\label{fig:fig1g}
\end{figure}
The \revisionSam{aim} of this work is to experimentally analyze the impact of buried structures at a variable depth on the PFM signal. To identify a suitable location with a domain transition, several overlapping PFM images of the wedge surface were taken and stitched together as shown in the supplemental Fig. S5(d). Here, a single frame containing one domain transition was chosen for the subsequent measurements. An example displaying the raw PFM amplitude as recorded by the instrument is shown in Fig. 3(a), where two distinct areas, corresponding to the domain configuration as shown in Fig. 1(a), can be identified. From such frames, data along lines orthogonal to the domain transition were extracted to generate data sets comparable to the theory as in Fig. 1(c). For an improved signal to noise ratio, the data in a larger area as highlighted in Fig. 3(a) is averaged via vertical binning. The resulting data is plotted in Fig. 3(b). Here, already the expected \revisionMatthias{$D^{-1}$} proportionality is readily seen in the inverted domain. Extracting the 90\%-depth from a partial fit of only the inverted domain yields a depth of $(553 \pm 20)$~nm, which readily shows that buried domains do impact the PFM signal. 

However, when comparing that signal shape with the theoretical prediction in Fig. 1 (c), quite a different shape is observed. Here, when inspecting a domain of sufficient depth with either a +z or -z domain orientation, one expects the PFM amplitude to reach the same value, while only the phase should change by 180$^{\circ}$. However, in the experiment shown in Fig. 3(a) and (b), it can be seen that the absolute PFM amplitude in the $+z$ domain is seemingly not converging towards the same level as in the $-z$ domain. This is a typical observation in many PFM investigations, \revisionMatthias{as the PFM signal is superimposed in phase-space by the intrinsic background of the experimental setup \cite{Jungk2006Quantitative}.}
Following the work of Soergel, \cite{pfm_soergel} this effect can be easily corrected for by analyzing both the PFM amplitude and phase channel, and the assumption that the PFM amplitude in the inverted domain should converge to the same level as in the non-inverted domain. Details on the correction method are discussed in the Supplement. Correcting the data yields the data set as displayed in Fig. 3(c). Fitting the model from Fig. 1(c) now yields the 90\%-depth to reach a value of $(1171 \pm 9)$ nm, which is almost twice as much as compared to the results from Fig. 3(b). Therefore, whenever the \revisionMatthias{PFM signal is superimposed by the intrinsic background}, the obtained depth information might be easily underestimated. 

Hence, when accurate quantitative data is required, a correction by a full mathematical \revisionMatthias{treatment} yields the best PFM results. Nevertheless, this is not always necessary. As shown in the supplement, if \revisionMatthias{the intrinsic background is sufficiently large} (more than twice the signal amplitude) the 90\% depth can be readily corrected by a flat multiplication factor of 19/9 (as shown in the supplement) for a buried domain. Here, 
\begin{equation}
 (553 \pm 20) \frac{19}{9}\:\:\text{nm} = (1170 \pm 40)\:\:\text{nm}     
\end{equation}
yields a depth of $(1170 \pm 40)$ nm, which is within accuracy limits identical to the results from fitting the corrected data set in Fig. 3(c). Such a correction should be sufficient for many cases where PFM is applied. Details on the correction method, as well as limits of using a flat multiplication factor can be found in the supplemental file. All subsequent data sets were corrected via this procedure.

\begin{figure*}
	\centering
	\includegraphics[width=0.9\linewidth]{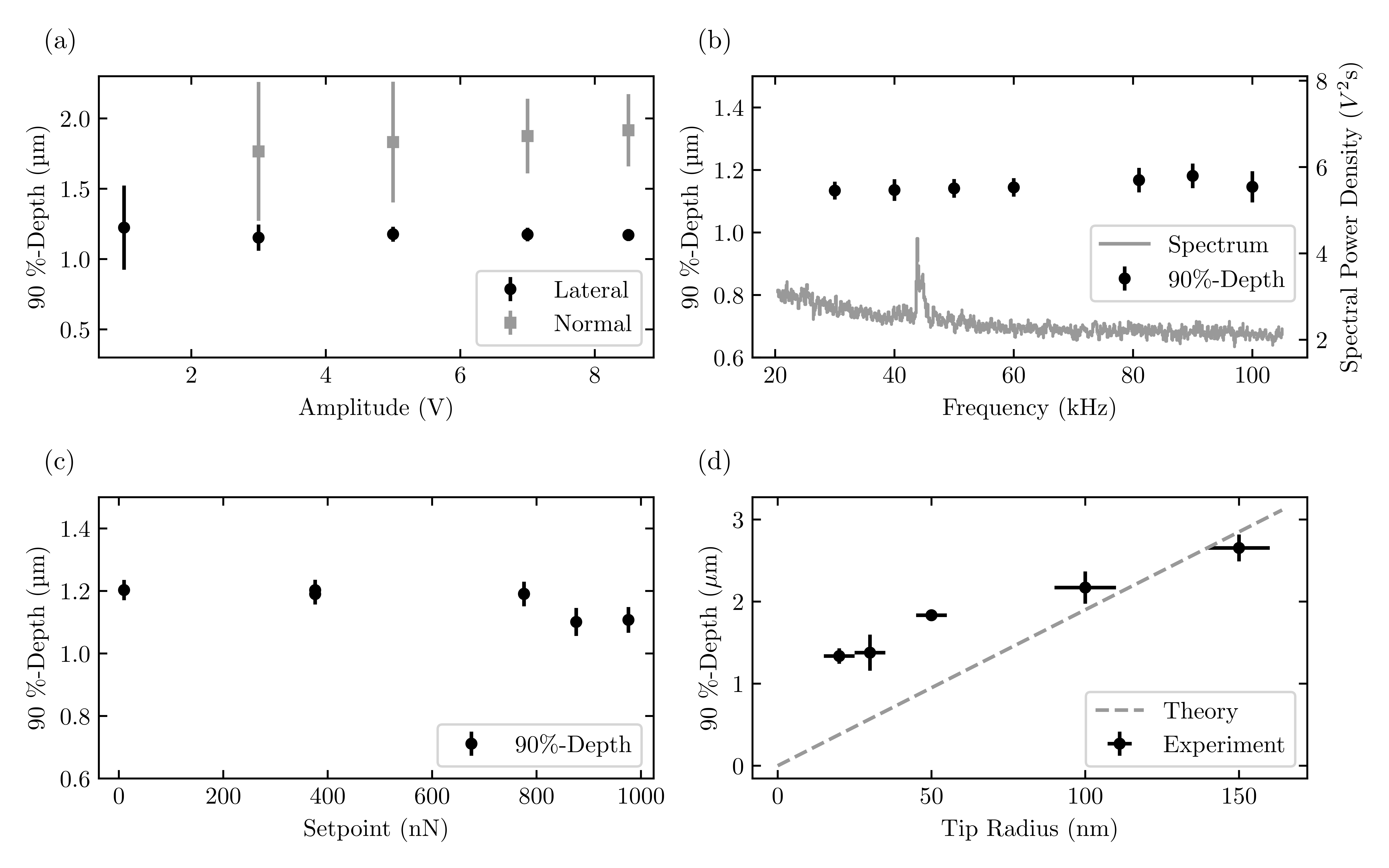}
	\caption{\revisionSam{Extracted values for the 90\% depth as a function of various parameters commonly tuned in PFM, namely (a) the applied ac oscillation voltage (for both the normal and lateral PFM signal); (b) the applied ac oscillation frequency; (c) the SFM tip force set point;
    and (d) the size of the SFM tip radius $r_{tip}$. The only dependence found was that on $r_{tip}$, as expected. For standard scanning parameters refer to the methods section.}}
	\label{fig:parameter_dependence}
\end{figure*}

\revisionSam{\subsection{Dependencies on PFM parameters}}
In order to explore the potential parameter space that influences the PFM probing depth, various scanning parameters were varied and compared as a function of the 90\% probing depth, as defined above. The results can be seen in Fig. \ref{fig:parameter_dependence}, where the applied PFM ac amplitude, frequency, tip contact force set-point, and tip radius were varied in (a-d), respectively. As expected from the theoretical calculations, the only dependence found was that on the PFM tip radius $r_{tip}$.

Fig. \ref{fig:parameter_dependence}(a) displays the 90\% depths extracted from normal %(out-of-plane)
and lateral (in-plane) PFM signals, as a function of PFM ac oscillation voltage. Here, also in the normal (out-of-plane) detection direction, a tiny depth-dependent signal was detected, which demonstrates that the direction and type of motion in these experiments can be neglected at first-order, as only the change in sign of the tensor element is relevant for uniaxial ferroelectrics like LNO. The normal PFM signal results from out-of-plane sample movement as well as cantilever buckling, while the lateral in-palne PFM signal is sensitive to torsion of the cantilever, only \cite{diss_haussmann}. The signal-to-noise ratio for the normal PFM signal is universally worse in these measurements than that of the lateral PFM signal, a fact that directly reflects into the error bars. For instance, in the case of the +1~V oscillation amplitude, it was in fact too noisy to extract a reliable PFM value, and was thus excluded. The fact that the lateral signal is much stronger as compared to the normal one is directly expected from the way the experiment has been set up. As discussed above, in the used geometry the vertical x-field emerging from the tip directly addresses the $d_{15}$ tensor element leading to a strong shear-motion in the xz-plane. The out-of-plane component detected in this experiment, however, may only stem from stray fields in the y or z direction, that do not fully average out. Moreover, the finite wedge angle might also provoke other tensor elements to be addressed, including elastic deformations of the crystal \cite{Kalinin2002,Kalinin2004}. 

Whilst the lateral signal extraction indicates a larger 90\% depth than the normal signal extraction, no dependencies were found as a function of magnitude of applied ac excitation voltage. The minor discrepancies seen in the difference between the fitted 90\% extracted from lateral and normal signals, is likely due to the reduced signal in the normal instance, given that the sample is poled in the lateral orientation.  

A similar lack of dependence of the PFM 90\% probing depth can be seen in the frequency dependence of the applied excitation voltage. Alongside the data extracted for the 90\% depth, in Fig. \ref{fig:parameter_dependence} (b) is a spectral power density spectrum indicating a tip-surface contact resonance of approximately 43~kHz. To avoid any parasitic effects induced by this contact resonance, measurements at this frequency were avoided. Extracted 90\% depth values were found to be in good agreement in the typical frequency range of 30~-~100~kHz. 

Due to the theoretical prediction that tip radius should directly change the probe depth of a given scan, it was also necessary to examine the effect of the cantilever force set point, given that this may challenge most directly the assumption of a SFM tip as a sphere due to indentation. Presented in Fig \ref{fig:parameter_dependence}(c), a consistent value was found for a loading up to $\sim$ 800~nN. Forces beyond that set point up to 1000~nN indicated a slightly lower extracted depth, however, yielding values that are still comparable to the `true' value. Note that such a huge tip loading is anyway beyond the typicall forces used to operate standard SFM and PFM.  

Finally, the effect of tip radius was examined, and presented in Fig \ref{fig:parameter_dependence}(d). A linear relationship between the tip radius and the 90\% depth is predicted. A variety of different tips with different $r_{tip}$ was applied, with both tip radii and error values taken from their respective data sheets. A good theory and experimental agreement can be seen for larger $r_{tip}$, with a less good agreement below $r_{tip}=$ 100~nm. Nonetheless, that experimental trend also indicates a linear dependence. This lack of agreement is likely due to the tip not being spherical for small $r_{tip}$, and other effects \cite{Kalinin2002,Kalinin2004}.

\revisionSam{\section{Conclusion}}
Despite being renowned as a surface sensitive technique, PFM inherently probes far deeper than the top layer of a sample. Analysis of periodically-poled domains in an uniaxial ferroelectric prepared to have constantly varying thicknesses of domains, has shown that the presence of domains buried below the surface are clearly visible beyond the 1~$\mu$m depth, more than 50 times larger as compared to a typical tip radius of 20~nm.

The probing depth was shown both mathematically and experimentally to depend entirely on the radius of curvature of the SFM tip, with all other typical PFM scanning parameters (applied ac volatage amplitude and frequency, tip force set point) to have no effect on the probing depth in these experiments. Despite the relatively simple analytical model, good quantitative agreement between theory and experiment could be achieved. Even the presence of an intrinsic background in measurement data can be quantitatively treated.

These results further motivate the use of PFM beyond simply imaging the 2D/3D ferroelectric's domain structure, but pushing PFM to become a quantitative, tomographic tool for exploring subsurface behaviors of polar nanostructures, as deep as $\sim 1\:\mu m$ into the (ferroelectric) material. These analyses reported here exhibit the caution required whenever inspecting thin-films or 2D (ferroelectric) materials below the 100-nm thickness.
%as are for instance periodically-poled thin-film lithium niobate samples as used for nonlinear and quantum optics \cite{SvenReitzig2021,Ruesing2019}; 
Since material applications of ferroelectrics typically need to be mounted on some substrate in order to carry out the PFM measurements, that substrate might equally affect the overall measured PFM signal, both in amplitude and frequency. \\

\section*{Acknowledgements}
The authors gratefully acknowledge financial support by the Deutsche Forschungsgemeinschaft (DFG) through projects FOR5044 (ID: 426703838), as well as the Würzburg-Dresden Cluster of Excellence on “Complexity and Topology in Quantum Matter” - ct.qmat (EXC 2147; ID 39085490). This work was supported by the Light Microscopy Facility, a Core Facility of  the CMCB Technology Platform at TU Dresden. Further, the authors thank Thomas Gemming and Dina Bieberstein for assistance in wafer dicing. The authors thank P. Mackwitz from Paderborn University for assistance with laser scanning microscopy.

\section*{Author Declarations}
\subsection*{Conflict of Interest}
The authors have no conflicts to disclose.
\subsection*{Availability of data}
The data that support the findings of this study are available
from the corresponding author upon reasonable request.

%\nocite{*}
\section*{REFERENCES}
\bibliography{aipsamp}% Produces the bibliography via BibTeX.

\end{document}